\documentclass[pra,twocolumn,floatfix,superscriptaddress,longbibliography,notitlepage, nofootinbib]{revtex4-1}
\pdfoutput=1

\usepackage[utf8]{inputenc}
\usepackage{braket}
\usepackage{amsmath}

\usepackage{dcolumn}   % needed for some tables
\usepackage{bm}        % for math
\usepackage{amssymb}
\usepackage{mathtools}
\usepackage{tikz}
\usepackage{qcircuit}
\usepackage{appendix}
\usepackage{hyperref}
\usepackage{subcaption}
\usepackage{amsmath,amsfonts,amssymb}
\usepackage{mathtools}
\usepackage{thmtools,thm-restate}
\usepackage{algorithm}

\usepackage{algpseudocode}

\newtheorem{theorem}{Theorem}
\newtheorem{lemma}{Lemma}

\usetikzlibrary{arrows.meta}

\def\BibTeX{{\rm B\kern-.05em{\sc i\kern-.025em b}\kern-.08em
    T\kern-.1667em\lower.7ex\hbox{E}\kern-.125emX}}

\begin{document}
\title{Quantum Algorithm for Estimating Largest  Eigenvalues}
\author{Nhat A. Nghiem}
\affiliation{Department of Physics and Astronomy, State University of New York at Stony Brook, Stony Brook, NY 11794-3800, USA}

\author{Tzu-Chieh Wei}
\affiliation{Department of Physics and Astronomy, State University of New York at Stony Brook, Stony Brook, NY 11794-3800, USA}
\affiliation{C. N. Yang Institute for Theoretical Physics, State University of New York at Stony Brook, Stony Brook, NY 11794-3840, USA}
\affiliation{Institute for Advanced Computational Science, State University of New York at Stony Brook, Stony Brook, NY 11794-5250, USA}
\begin{abstract}

  Scientific computation involving numerical methods relies heavily on the manipulation  of large matrices, including solving linear equations and finding eigenvalues and eigenvectors. Quantum algorithms have been developed to advance these computational tasks, and some have been shown to provide substantial speedup, such as factoring a large integer and   solving linear equations. In this work, we leverage the techniques used in the Harrow-Hassidim-Llyod (HHL) algorithm for linear systems, the classical power, and the Krylov subsapce method to devise a simple quantum algorithm for estimating the largest eigenvalues in magnitude of a Hermitian matrix. Our quantum algorithm offers significant speedup with respect to the size of a given matrix over classical algorithms that solve the same problem. 
   
\end{abstract}
\maketitle

\section{Introduction}
Quantum computation  has undergone a long way since its  earliest conceptions~\cite{feynman2018simulating, deutsch1985quantum}. Many quantum algorithms have been proposed to tackle computational problems, ranging from searching~\cite{grover1996fast}, factoring~\cite{shor1999polynomial}, simulating quantum systems~\cite{childs2010quantum, feynman2018simulating, berry2007efficient},  solving linear systems~\cite{harrow2009quantum, childs2017quantum}, machine learning and data science~\cite{lloyd2013quantum, lloyd2014quantum, lloyd2016quantum, lloyd2020quantum, schuld2018supervised, mitarai2018quantum}, and to problems in computational topology and geometry~\cite{lloyd2013quantum, nghiem2022constant}. One of the most versatile results in the above is the quantum algorithm--the so-called HHL algorithm--for solving linear systems,  proposed by Harrow, Hassidim, and Lloyd~\cite{harrow2009quantum} (as well as its improvement~\cite{clader2013preconditioned,childs2017quantum,wossnig2018quantum}), as linear systems play a central role in many engineering and  science domains. Aside from its potential exponential speedup, another important result from Ref.~\cite{harrow2009quantum} is the BQP-completeness of the matrix inversion, which is a more general statement of the complexity regarding solving linear systems. The HHL algorithm has also been improved in Ref.~\cite{childs2017quantum}, which yields an exponential improvement in the error dependence. While quantum advantage has been  theoretically claimed, experimental realizations for  practical applications  remain to be seen.\\

In this article,  motivated by the results in Refs.~\cite{harrow2009quantum, childs2017quantum}, we consider the following problem and its generalization: \\

\textit{\textbf{Problem:}  Given a sparse Hermitian matrix $A$ of size $n \times n$ with bounded eigenvalues: 
\begin{align}
    0 < |\lambda_1| \leq |\lambda_2| \leq \cdots \leq |\lambda_n| < 1,
    \label{eqn: boundedrange}
\end{align}
Estimate its largest (in magnitude) eigenvalue $\lambda_n$.} \\

We note that even if the range of eigenvalues is not within this simple range, then by a trivial scaling, we can reduce the range of eigenvalues of all matrices to $(0,1)$, similar to that used in Ref.~\cite{harrow2009quantum}. Therefore, in general, it is safe to work with this range of eigenvalues without losing generality. Below, we construct a quantum algorithm that, given access to entries of a sparse Hermitian matrix $A$ in a  manner similar to that in Ref.~\cite{harrow2009quantum}, can estimate its largest eigenvalue in magnitude with a running time that is  logarithmic with respect to the size of  $A$. The main tool of our algorithm is built upon the HHL-like quantum linear system~\cite{harrow2009quantum}, or more precisely, the modified version of the HHL algorithm from Ref.~\cite{wiebe2012quantum}, the Hadamard test~\cite{nielsen2002quantum} and in particular, the power method. As we will further discuss, with respect to  dimension $n$ of matrix $A$, our algorithm provides quadratic speedup against, e.g., the classical power method in the worst case. Under some favorable conditions regarding initialization, exponential speedup with respect to dimension $n$ could be achieved.  In the following, we shall also generalize our algorithm to the case of the eigenvalue of the lowest magnitude and of the largest and lowest real values, as well as a hybrid quantum-classical Krylov subspace generalization.

 The remaining structure of this paper is as follows.  First, in Sec.~\ref{sec: classicalalgorithm}, we review the classical algorithm by the \textit{power method}, which can estimate the largest absolute eigenvalue of some  matrix $A$, which in general, does not have to be sparse or well-conditioned. In Sec.~\ref{sec: qa}, we present a corresponding quantum algorithm that solves the largest eigenvalue problem mentioned above, and it is a direct translation of the classical power method into the quantum setting. Our main result is stated as Theorem~\ref{thm: quantumrunningtime}. The advantage over the classical algorithm is discussed subsequently in Sec.~\ref{sec: advantage}. In Sec.~\ref{sec: discussion}, we then discuss a few possible extensions of our quantum algorithm, including the non-sparse case and how to modify our algorithm for the \textit{minimum eigenvalue}, as well as an extension to a few largest eigenvalues in magnitude. We conclude in Sec.~\ref{sec:conclusions}. Additionally, in Appendix~\ref{sec: erroranalysis}, we provide an analysis of the power method, showing that the number of iterations is efficiently low in order to reach the convergence of the power method. In Appendix~\ref{sec: HHL}, we revise the original HHL approach and show how it is adapted to our case, and in Appendix~\ref{app:Improved}, we show a simplified version of the multiplication application of the matrix, reducing the time complexity. In Appendix~\ref{sec: krylov}, we give more detail on how our algorithm can be extended to build the Krylov subspace and obtain the largest few eigenvalues.

\section{Classical Algorithm}
\label{sec: classicalalgorithm}

The classical approach to tackle the above problem is called the \textit{power method} \cite{chu1993multivariate}, which is a very standard approach \cite{parlett1982estimating,o1979estimating}, and literally can be found in any textbook on numerical methods. The main procedure is as follows:  \\

$\bullet$ First initialize some random  vector $x_0 \in C^n$. \\

$\bullet$ Iterate the matrix application process $k$ times and form the following sequence: 
\begin{align*}
    x_1 = A  x_0, \\
    x_2 = A  x_1 = A^2  x_0, \\
    x_3 = A x_2 = A^3 x_0, \\
    \vdots \\
    x_k = A^k x_0.
\end{align*}

$\bullet$ Compute the approximated largest eigenvalue $\bar{\lambda}_n$: 
\begin{align*}
    \bar{\lambda}_n \equiv\frac{Ax_k^* \cdot x_k}{x_k^* \cdot x_k},
   % \label{eqn: classicaleigen}
\end{align*}
where $(\cdot)$ denotes the inner product, and $*$ denotes the complex conjugate. We remind that, since $A$ is Hermitian, its eigenvalues are always real.  \\

The following lemma shows that $x_k$ can approximate the eigenvector $E_n$ with eigenvalue $\lambda_n$. 
\begin{lemma}
\label{lemma: approx}
Let $x_k$ be defined as above. Assuming the largest absolute eigenvalue $\lambda_n$ (with corresponding eigenvector $E_n$) is unique, i.e., $|\lambda_i| < |\lambda_n|$ for all $1< i <n$. Then 
\begin{align*}
    x_k \approx \alpha \cdot E_n,
\end{align*}
where $\alpha$ is a some number that depends on $k$--the number of iterations.
\end{lemma}
\textit{Proof:} Since $A$ is Hermitian, its eigenvectors $\{E_n\}_1^N$ can be made orthonormal if degenerate. We can choose them to be the basis of $A$. Therefore, the initial random vector $x_0$ can be expressed as:
\begin{align}
    x_0 = c_1E_1 + c_2E_2 + \cdots + c_nE_n.
    \label{eqn: initialx0}
\end{align}
After the first iteration, we have
\begin{align}
    x_1 &= Ax_0 \\
    &= A \big(c_1E_1 + c_2E_2 + \cdots + c_nE_n \big) \\
    %&= c_1 A E_1 + c_2 A E_2 + \cdots + c_n A E_n \\
    &= c_1 \lambda_1E_1 + c_2 \lambda_2E_2 + \cdots + c_n \lambda_n E_n.
\end{align}
Proceeding similarly, after $k$ iterations, we have
\begin{align}
    x_k 
    %&= c_1 \lambda_1^k E_1 + c_2 \lambda_2^k E_2 + \cdots + c_n \lambda_n^k E_n \\
    %&
    = \lambda_n^k \Big( c_1\frac{\lambda_1^k}{\lambda_n^k} E_1 + c_2 \frac{\lambda_2^k}{\lambda_n^k} E_2 + \cdots  + c_n E_n \Big).
    \label{eqn: xk}
\end{align}
We have assumed that the largest eigenvalue in magnitude is non-degenerate,
$$ 0 \leq |\lambda_1| \leq |\lambda_2| \leq \cdots < |\lambda_n|. $$
Therefore, 
%each of the fraction
%$$ \frac{|\lambda_i|}{|\lambda_n|} < 1,  $$
%for all $i<n$. It is straightforward to see from the above property that
\begin{align}
    \frac{\lambda_i^k}{\lambda_n^k} \rightarrow 0, \ \mbox{as} \, k \rightarrow \infty.
\end{align}
%as $k \rightarrow \infty$. 
Hence, when $k$ is sufficiently large, we have:
$$  x_k \approx \lambda_n^k c_nE_n \sim E_n. $$ 
The proof is then completed. $\blacksquare$ \\

From here, we can see the factor $\alpha$ in Lemma.~\ref{lemma: approx} is 
$$  \alpha = \lambda_n^k c_n.$$
Now we try to compute:
$$ \frac{Ax_k^* \cdot x_k}{x_k^* \cdot x_k}. $$
Since $x_k \approx \alpha E_n$, then $Ax_k \approx \lambda_n x_k$. The inner product is then:
$$ Ax^*_k\cdot x_k \approx \lambda_n ( x^*_k \cdot x_k), $$
which directly leads to 
\begin{align}
     \frac{Ax^*_k \cdot x_k}{x^*_k \cdot x_k} \approx \lambda_n.
     \label{eqn: classicalratio}
\end{align}

In Appendix~\ref{sec: erroranalysis}, we provide a careful analysis of the above approximation. It turns out that the number of iteration $k$ grows  only logarithmically with the inverse of the error, thus showing the efficiency of the power method. We note further that if the initially randomized vector $x_0$ is orthogonal to the eigenvector $E_n$ corresponding to the largest eigenvalue, then the power method will not work as the consecutive application of matrix $A$ would not `drive' $x_0$ closer to $E_n$. In this case, a simple trick is to run the power method multiple times with other randomly chosen $x_0$ and compare the results. Since $x_0$ is randomized, then only with a small probability that $x_0$ would completely lie in the subspace orthogonal to $E_n$.  Even if $c_n$ is much smaller than other coefficients, the iteration increases its magnitude relative to others exponentially fast. To elaborate on this, we do a simple observation: after $k$ iterations, the coefficient corresponding to $E_n$ would be $c_n \lambda_n^k$. If we want $c_n \lambda_n^k \approx \mathcal{O}(1)$, then it is sufficient for $k = \mathcal{O}( \log(1/c_n) )$. More important is the relative magnitude, e.g., $c_n/c_{n-1}$, which  we assume  is much less than unity. Then  the comparison is such that $(c_n/c_{n-1}) \lambda_n^k /\lambda_{n-1}^k \gtrsim1$. So $k$ will be chosen to be $$\mathcal{O}( \lambda_{n-1}/\lambda_n\log(c_{n-1}/c_n) ).$$

The power method is improved by the Lanczos algorithm~\cite{chu1993multivariate}, which essentially builds up the Krylov subspace spanned by $[x_0, Ax_0, A^2 x_0, \dots, A^k x_0]$. By reducing $A$ to this subspace, one can solve 
 the largest few eigenvalues. We note that in this way the non-degeneracy constraint above (in Lemma~\ref{lemma: approx}) can be lifted. \\

\section{Quantum Algorithm}
\label{sec: qa}
In this section, we show how to translate the above classical power method into the quantum setting and find the largest eigenvalue. We begin by noting that, per the power method, the approximated largest eigenvalue is given as:
\begin{align*}
    \label{eq:lambdan}\bar{\lambda}_n \equiv\frac{Ax_k^* \cdot x_k}{x_k^* \cdot x_k},
\end{align*}
where $x_k = A^k x_0$. The denominator $x_k^* \cdot x_k$ is exactly the square norm of $x_k$, i.e, $|x_k|^2$. We further remark that 
\begin{align}
    \frac{x_k}{ |x_k|} \equiv \ket{x_k}
\end{align}
is exactly the normalized quantum state corresponding to $x_k$. Therefore, the above $ \bar{\lambda}_n $ can be written as:
\begin{align}
     \bar{\lambda}_n  = \bra{x_k} A \ket{x_k}.
\end{align}
Now we aim to estimate the largest eigenvalue using the above formula. 

\subsection{Main Procedure}
We first state the following lemma, which is based on the HHL algorithm~\cite{harrow2009quantum} and its adaptive version~\cite{wiebe2014quantum}. In Appendix~\ref{app: HHL}, we review these methods.                      
\begin{lemma}
\label{lemma: matrixapp}
    Given access to an $s$-sparse, Hermitian matrix $A$ of size $n \times n$. Then there exists a unitary $U_A$ that performs the following transformation on a given initial state $\ket{0}\ket{x_0}$:
    \begin{align}
    \label{eq:UA}
        U_A \ket{0}\ket{x_0} =  C\ket{0}  \Tilde{A} \ket{x_0} + \ket{1} \ket{ Garbage },
    \end{align}
where $\Tilde{A} \ket{x_0}$ refers to an approximated version of $A\ket{x_0}$, and $C$ is some constant that guarantees the normalization condition (see Sec. \ref{sec: HHL}), e.g, $C \lambda_{max} \leq 1$ where $\lambda_{max}$ is the largest eigenvalue of A. The running time of $U_A$ is: 
    \begin{align*}
        \mathcal{O}\Big( \log (n) s \kappa / \epsilon  \Big), 
    \end{align*}
    where $\kappa$ is the conditional number of A, and $\epsilon$ is the error tolerance. 
\end{lemma}

Throughout this work, we will make the following abuse of notation: we will use $A\ket{x_0}$ to denote the approximated version and assume that the error due to truncation is implicit, i.e, $A\ket{x_0}$ refers exactly to $\Tilde{A} \ket{x_0}$ in \ref{lemma: matrixapp}. Further, as we have assumed from the beginning that the range of eigenvalues is $(0,1)$, therefore, the constant $C$ could be set to 1, as similar to \cite{wiebe2012quantum}. Otherwise, if the range of eigenvalues is arbitrary, either we rescale the matrix as we mentioned previously, or there will be a factor $C \approx \mathcal{O}(1/\lambda_{max})$ enters the analysis that we would discuss subsequently. Though, it would not affect the overall running time, as it will be absorbed through the conditional number $\kappa$. Throughout, we set $C=1$ for simplicity.\\

Using the above lemma, we can see that, given the initially randomized $\ket{x_0}$ (we may also use $x_0$ to denote, for notation simplicity), if we wish to apply $A$ to it $k$ times, we simply need to repeat the above procedure in~(\ref{eq:UA}) $k$ times, followed by a measurement to post-select the first part. However, it is very interesting to note that such $k$-iterations  can be done faster quantumly and  actually in one go (and thus $k$ is just a parameter, not the steps of iterations). The rough idea is that, based on the HHL algorithm~\cite{harrow2009quantum}, we aim to extract the eigenvalues of $A$, then rotate the ancilla conditioned on the phase register. Since the eigenvalues of $A$ are stored after the quantum phase estimation step in implementing~(\ref{eq:UA}), then we can apply arbitrary arithmetic operations on them directly. Here we state the main tool that we would use subsequently:
\begin{lemma}
    In a similar assumption as in Lemma~\ref{lemma: matrixapp}, the following unitary can be achieved: 
    \begin{align}
        U_{A^k} \ket{0}\ket{x_0} = \ket{0} A^k \ket{x_0} + \ket{1} \ket{Garbage},
        \label{eqn: kiteration}
    \end{align}
 with the running time: 
\begin{align*}
        \mathcal{O}\Big( \log (n) s \kappa / \epsilon \Big). 
    \end{align*}
    \label{lemma: improvedmatrix}
\end{lemma}
More details will be discussed in Appendix~\ref{app:Improved}, including the proof of the above lemma. Note that in general, we have suppressed a factor $C^k$ in the normalization; see also the previous Lemma.  \\

Now we proceed to the final goal, which is estimating the largest eigenvalue. Note that we denote $x_k \equiv A^k \ket{x_0}$; therefore, if we measure the first ancilla in the state~(\ref{eqn: kiteration}) and observe $\ket{0}$, we will obtain the desired state $\ket{x_k}$ (renormalization of the state is naturally taken care of). The probability of measuring $\ket{0}$ in the first register at the $k$-step is, from Eq.~(\ref{eqn: kiteration}),
\begin{align}
    p_0(k) = | A^k\ket{x_0}|^2 = \sum_{i=1}^n |c_i|^2 \lambda_i^{2k}.
\end{align}

As we discussed, there should be a factor C in the above probability in a more general setting, but we set $C$ to 1. Remarkably, what we would discuss here still works even with the most generality. Since we have assumed that for all $i$, $\lambda_i > 1/\kappa$, therefore, $p_0$ is lower bounded by $1/\kappa^{2k}$. The number of repetition may grow as large as $\mathcal{O}(\kappa^{2k})$, which can be improved to $\mathcal{O}(\kappa^{k})$ by using amplitude amplification \cite{brassard2002quantum}, if we wish to obtain $\ket{0}$. 

Suppose we have obtained the desired state $\ket{x_k}$, we now present the so-called Hadamard procedure to estimate the quantity $\bar{\lambda}_n  = \bra{x_k} A \ket{x_k}$. \\

\smallskip
\noindent\textbf{Hadamard Procedure:} 
We begin with $\ket{+}\ket{0}\ket{x_k}$, which can be prepared from $\ket{0}\ket{0}\ket{x_k}$ and apply Hadamard gate on the first register. We then apply a conditional $U_A$ (from Lemma \ref{lemma: matrixapp}) depending on the first qubit to obtain:
\begin{align}
    \frac{1}{\sqrt{2}} \Big(  \ket{0}(  \ket{0} A\ket{x_k} + \ket{1}\ket{Garbage}  )  + \ket{1}\ket{0}\ket{x_k}    \Big)
\end{align}
Measuring the first qubit in the $\ket{+/-}$ (Pauli X) basis gives the real part of the overlap, e.g,  $\bra{x_k} A \ket{x_k} $. Additionally, measuring the first qubit in the $\ket{+i/-i}$ (Pauli Y) basis gives the imaginary part of the overlap. We highlight that in the general setting (see lemma \ref{lemma: matrixapp}), there is a factor $C$. Therefore, what we estimate here is up to such a known normalization factor.

We now analyze the algorithm as a whole and state its scaling time depending on dimension n, sparsity s, conditional number $\kappa$, and error desire $\epsilon$.

\subsection{Time Complexity}
Recall that our quantum algorithm is built upon the classical power method straightforwardly. For a fixed number $k$ of iteration steps, we first perform the multiplication of initially randomized vector $\ket{x_0}$ with given matrix $A^k$ (see lemma \ref{lemma: matrixapp} and \ref{lemma: improvedmatrix}). Measurement and post-selection steps are done to obtain the (normalized) state $\ket{x_k}$. The largest eigenvalue of $A$ can be estimated via the combination of  $\ket{x_k}$, the ancilla,  another round of matrix application $A$ (more precisely, a  conditional version of $A$), and a final Hadamard test.  

Since several main subroutines are involved, we remark that there can be multiple sources of error in our quantum algorithm. The matrix application step (\ref{lemma: improvedmatrix}) introduces an error $\epsilon$, i.e., all the states $\ket{x}$ that we obtain after every post-measurement step is, in fact, $\epsilon$-close to the ideal state $\ket{x}_{ideal}$ (note we have  abused the notation as mentioned previously). We further remark that the Hadamard test requires $\mathcal{O}(1/\epsilon^2)$ repetitions in order to estimate the overlaps to some additive accuracy $\epsilon$.

The running time of our algorithm accumulates from the matrix application (\ref{lemma: improvedmatrix}), measurement plus post-selection, and the Hadamard test. The most consuming step is the measurement and post-selecting $\ket{0}$ on the first qubit. As we mentioned, achieving the desired state takes $\mathcal{O}(\kappa^k)$ repetitions. We remind that the power method has a random factor: the initialization of the random vector $x_0$. In Appendix~\ref{sec: erroranalysis}, we provide careful analysis and discussion of the power method from the classical perspective. It is shown that, with high probability, for the power method to achieve an error $\mathcal{O}(\epsilon)$, the value of $k$ needs to be of $\Omega( \log(1/\epsilon) + \log \sqrt{n})$, which means that the measuring/post-selecting step will take time $\mathcal{O}( \kappa^k ) = \mathcal{O}( \kappa \sqrt{n}/\epsilon)$. However, if the initially randomized $x_0$ is  already close to the largest eigenvector of $A$, then as we discuss in detail in Appendix~\ref{sec: erroranalysis}, the running time can be reduced to $\mathcal{O}(\kappa/\epsilon)$.

All in all, the main result of our work is stated in the following theorem. 
\begin{theorem}
\label{thm: quantumrunningtime}
    Given access to a sparse Hermitian matrix A of size $n \times n$. By employing the power method with an initial vector $x_0 = \sum_{i=1}^n c_i E_i$ (as defined in \ref{eqn: initialx0}), its largest eigenvalue (in magnitude) $\lambda_{max}$ can be estimated to some accuracy $\epsilon$ in time
\begin{align}
    \mathcal{O}\Big(   \frac{\log (n) \sqrt{n} s \kappa^2}{ \epsilon^4}   \Big).
\end{align}

If for all $i \neq n$, $c_n/c_i \geq \sqrt{n-1}$, the running time can be reduced to
$$ \mathcal{O}\Big(   \frac{\log (n) s \kappa^2}{ \epsilon^4}   \Big).$$
\end{theorem}

We note that $1/\epsilon^4$ comes from the combination of the phase estimation, the Hadamard test, and the error induced by the power method itself.
The above result suggests  a critical role of the conditional number $\kappa$ in our algorithm, as the running time depends on it \textit{polynomially}. While we do not expect $\kappa$ to grow  faster than polylogarithmic of $n$, the problem regarding the conditional number has been raised and resolved in Ref.~\cite{clader2013preconditioned} by using a preconditioning technique. The same technique can be used in our case, as one only needs to replace the original matrix, say, $A$ by a new matrix $A'$ with a bounded conditional number~\cite{clader2013preconditioned}, and then proceed with the same algorithm as we have outlined above. Whether or not one can reduce the dependence  on the conditional number $\kappa$ directly in the quantum matrix multiplication is an interesting open question. We also emphasize that regarding the error $\epsilon$ dependence, the most time-consuming factor is the Hadamard test, which takes $\mathcal{O}(1/\epsilon^2)$ trials. Thereby, it is also very worthy of finding better solutions to  estimating the inner product, e.g., whether  one could improve the running time to $\mathcal{O}(1/\epsilon)$ or even $\mathcal{O}(\log 1/\epsilon)$.  \\

As a final remark, in addition to the maximum eigenvalue, our procedure also yields a corresponding approximate eigenvector, which can be used for further processing, such as computing expectation values of observables. \\

\subsection{Advantage Over Classical Algorithm}
\label{sec: advantage}

Now we discuss the advantage that our quantum algorithm can yield in this problem. We remind readers that the quantum algorithm that we outlined above is essentially a translation of the classical power method to the quantum setting. While the running time depends most critically on the conditional number $\kappa$ (for which we just  pointed out a solution from Ref.~\cite{clader2013preconditioned} to reduce the complexity), the logarithmic dependence on dimension $n$ is quite remarkable.

\smallskip\noindent {\bf 
 Classical algorithm}.  To the best of our knowledge, the best classical algorithm for sparse matrix multiplication, 
 such as described in Ref.~\cite{yuster2005fast}, achieves the running time $\mathcal{O}(s^{0.7} n^{1.2} + n^{2+O(1)} )$. Even when the sparsity $s$ is not large, i.e., $s \ll n$, multiplying two sparse matrices still takes time $\mathcal{O}(n^{2}) $, which is polynomial in $n$. After the multiplication step, a further $\mathcal{O}(n)$ time is required to estimate the inner products; see the numerator and denominator of formula~(\ref{eqn: classicalratio}). However, we observe that while the matrix $A$ is $s$-sparse, the multiplication of $A$ by itself, i.e., $A^2$ is not necessarily sparse. In the classical power method, we are required to multiply $A$ by itself $k$ times. Therefore, after the first step of multiplication, we can no longer  use sparse multiplication~\cite{yuster2005fast}, but the traditional matrix multiplication method, e.g., Strassen's algorithm, which has time complexity $\mathcal{O}(n^3)$. This approach is apparently costly when $n$ is large. 
 
 Alternatively, one can choose to first perform $A x_0$ (which takes time $\mathcal{O}(n s)$ when $A$ is $s$-sparse)  then follow with another $(k-1)$ rounds of multiplication by $A$. The running time of this approach would be $\mathcal{O}(s n k)$, which is better than the previously described method. As we can see, the classical method is more straightforward than the quantum procedure that we outlined previously. \\

\smallskip\noindent {\bf 
 Comparison}. Comparing between classical algorithm (with running time $\mathcal{O}(snk)$) and quantum algorithm (see Thm~\ref{thm: quantumrunningtime}), there is quadratic speedup for the latter on dimension $n$ in general cases. The speedup regarding $n$ can be improved to exponential if the initial seed $x_0$ satisfies the closeness to the largest eigenspace of $A$, as we emphasized in Thm~\ref{thm: quantumrunningtime} and is elaborated in Appendix~\ref{sec: erroranalysis}.  

\section{Some Extensions and Relevant Works}
\label{sec: discussion}
The problem that we solve in this paper is part of the \textit{matrix and eigenvalue problem}.  
Despite seeming simple, such a problem has far-reaching consequences, in both theory and application. In pure and applied mathematics, one of the very difficult problems concerns the distribution of eigenvalues of random matrices~\cite{tao2011random}. In physics, a system is characterized by the Hamiltonian, which is represented by a Hermitian matrix $H$. The ground state is the particular state that corresponds to the eigenstate of $H$ having the smallest algebraic eigenvalue, to which our algorithm can be modified. 

Until now, we have assumed  that the matrix $A$ is Hermitian, but the extension to non-Hermitian matrices is straightforward as discussed in Ref.~\cite{harrow2009quantum} using a new matrix which places $A$ and $A^\dagger$ in its two off-diagonal blocks,
\begin{equation}
  \tilde{A}=\begin{pmatrix}
    0 & A \\
    A^\dagger & 0
    \end{pmatrix},
\end{equation}
and then proceeding with the new Hermitian matrix $\tilde{A}$. Alternatively, if one can find a way to simulate $\exp(-iA^\dagger A t)$ then, in principle, our algorithm can still work, as the eigenvalues of $A^T A$ are the square of singular values of $A$. In both cases, one can estimate the largest singular value of a given matrix $A$. Below, we  will discuss the further extensions of our work and point out some overlaps/connections to previous works. 

\subsection{Non-Sparse Matrix}
We recall that we have assumed  the sparsity of the matrix $A$. At the core of our algorithm is the application of $A$, which requires the ability to simulate $\exp(-iAt)$ to be used by the quantum phase estimation. As in \cite{harrow2009quantum}, such simulation is doable if $A$ is sparse and row-computable, thanks to the method in~\cite{berry2007efficient}. However, even if the Hermitian matrix $A$ is non-sparse, positive, and it can be decomposed as 
$$ A = B^\dagger B,$$
 and simulating the time evolution is still possible, provided there exists a way to prepare the following state
$$ C\sum_i |B_i| \ket{i}\ket{B_i}, $$
where $C$ is an overall constant, $B_i$ refers to the vector columns of $B$, and $|B_i|$ is the $l_2$-norm of such vector. We can thus employ  the method proposed in Ref.~\cite{lloyd2014quantum} to simulate $\exp(-iAt)$ by  the \textit{density matrix exponentiation}. With such an ability to simulate $\exp(-iAt)$, quantum phase estimation of eigenvalues of $A$ is thus possible. Hence, we can execute our quantum algorithm  as above to find the maximum eigenvalue of $A$ (possibly with some scaling/or normalization factor).  

In fact, a similar problem was encountered in Ref.~\cite{lloyd2014quantum}, and the authors proposed to use density matrix exponentiation combined with quantum phase estimation to solve the problem called principle component analysis. Roughly speaking, they aimed to find a few largest eigenvalues (and corresponding eigenvectors) of the so-called covariant matrix, represented by some density matrix $\sum$, assuming that there exists a way to prepare $\sum$. Once the exponentiation $\exp(-i\sum t)$ is obtained, running quantum phase estimation algorithm (QPE)~\cite{kitaev1995quantum, brassard2002quantum}, with the density matrix $\sum$ itself also being prepared as an initial state, allows us to reveal the highest value of the spectra by sampling. The method works well when a few top eigenvalues have much higher values than the remaining ones (which is why they are called principle components). In our problem, we remark that once $\exp(-iAt)$ is obtained through a known procedure, QPE can be used directly to reveal the maximum eigenvalue.  We can, for example, run QPE with $\exp(-iA\cdot 2\pi)$ as the main unitary and the maximally mixed state $I/n$ as the input state. We then approximately obtain the following:
\begin{align}
    \frac{1}{n}\Big( \sum_{i=1}^n \ket{\Tilde{\lambda_i}}\bra{\Tilde{\lambda_i}} \otimes \ket{E_i}\bra{E_i}   \Big).
\end{align}
Sampling from the above state will, in fact, reveal full spectra, including the maximum eigenvalue. However, as simple as it may seem, this is not  efficient as the probability of obtaining different eigenvalues is the same ($=1/n)$ and hence, it requires as much as $\mathcal{O}(n)$ time to obtain the maximum one. If we apply the controlled rotation to ancilla in order to multiply the eigenvalues to the corresponding components, then this essentially reduces to the quantum principle component analysis~\cite{lloyd2014quantum}, discussed above. 

An alternative solution to non-sparse $A$ is a direct result of~\cite{wossnig2018quantum}, where the authors consider the problem of solving a dense linear system. Given the particular data structure of $A$ and an oracle access to $A$ (see Lemma 1 in~\cite{wossnig2018quantum, kerenidis2016quantum}), they outline a somewhat similar strategy to Ref.~\cite{harrow2009quantum} in order to solve the linear system. The goal is to coherently extract the singular values of $A$ (which are eigenvalues in the Hermitian case) and rotate the ancilla (so as to multiply the inverse of the eigenvalue). Similarly, using  an adaptive method introduced in~\cite{wiebe2012quantum}, we can multiply the eigenvalues instead  to apply the matrix $A$ instead of $A^{-1}$. A potential caveat of this approach is that the exponential speedup (with respect to the size of the matrix) might not be obtained, but only a quadratic speedup instead, as discussed concretely in~\cite{wossnig2018quantum}.

\subsection{Finding The Minimum Eigenvalue}
For a Hermitian matrix, the conditional number of a matrix $\kappa$ is defined as:
$$ \kappa = \frac{\lambda_{max}}{\lambda_{min}}. $$
Note that this formula is only exact for a Hermitian matrix. In general, singular values are taken into account instead of eigenvalues.
In principle, if the conditional number of a matrix is known, once the maximum eigenvalue of a matrix is revealed by the above algorithm, then the minimum eigenvalue can be estimated. However, in cases where the conditional number is not exactly known (we might, for example, know its upper bound), then it is hard to have a good estimate of the minimum eigenvalue. Here, we aim to tackle such a problem using the algorithm that we have developed above. Let $\{\lambda_i\}_{1}^n$ be eigenvalues of $A$. We assume that $A$ is invertible, i.e., does not contain zero eigenvalues (otherwise,  one can use a shifted matrix $A - cI$ to avoid this), then the eigenvalues of $A^{-1}$ is 
$$ \Big\{ \frac{1}{\lambda_i}\Big\}_{i=1}^n. $$

If $\lambda_{max}$ is the maximum eigenvalue of $A$, then $1/\lambda_{\max}$ is the minimum eigenvalue of $A^{-1}$, and similarly, if $\lambda_{\min}$ is the minimum eigenvalue of $A$, then $1/\lambda_{\min}$ is the maximum eigenvalue of $A^{-1}$. So instead of $A$, we simply apply $A^{-1}$ as in the original HHL, then our above procedure  with this modification can find the maximum eigenvalue of $A^{-1}$, from which we can find the minimum eigenvalue of $A$. \\

\subsection{Shifting by an Identity Matrix to Compute Ground-State Energy}
As we have mentioned above, physical systems are generally characterized by a Hamiltonian $H$, which is Hermitian. Thus, analyzing its spectrum, including finding the ground state (and possibly a few low-lying excited states) and its corresponding energy, is one of the most important problems in physics. In our algorithm, as an alternative solution to what we have described in  the previous subsection, we can also consider a shift by an identity matrix $A-cI$ that shifts the spectrum by a constant $c$. Thus one can compute the $\max_i |\lambda_i-c|$. Shifting has been a useful technique, for example, in finding the excited states and their spectra in many-body localization~\cite{yu2017finding}. The energy level statistics in potential many-body localized systems can be gained from applying such a technique and  be sped up in principle by our quantum algorithm.  

If one applies a large enough shift so that all the eigenvalues are negative, then the one corresponding to the largest magnitude gives the lowest eigenvalue of $A$ and the procedure also yields its eigenvector, i.e., the ground state.  
In a similar way, if one adds a sufficiently large positive shift, so all the eigenvalues are positive, then one can find the largest algebraic eigenvalues. 

If one wishes to reveal the ground-state energy, then our quantum power method achieves a highly efficient running time with respect to the overlaps between the initial randomized state and ground state. Let $\eta$ denote such overlaps. We have mentioned in Sec.~\ref{sec: classicalalgorithm} that it would take $\mathcal{O}(\log (1/\eta))$ to `drive' the initial state $\ket{x_0}$ close to the ground state $\ket{E_1}$, and hence we would be able to estimate the ground-state energy with  further  additional time that is independent of $\eta$. In Ref.~\cite{wang2022quantum}, the authors proposed a quantum algorithm for estimating ground state energy with running time $\mathcal{O}(1/\eta^2)$. Hence, with respect to $\eta$, our quantum power method has a much better scaling.  In Ref.~\cite{lin2022Heisenberg}, to achieve Heisenberg-limited precision scaling, their algorithm needs a total  evolution time to scale with $\mathcal{O}(poly(1/\eta^2))$, despite their maximal coherent runtime at each run scaling as $\mathcal{O}(polylog(1/\eta^2))$.

Recall that our quantum algorithm is essentially a direct translation of the classical power method, where the aim is to consecutively apply the  matrix $A$ to some initially random state. Another approach that also shares similarities with the classical power method is quantum imaginary time evolution (QITE) \cite{motta2020determining}. In this approach, one wishes to find the ground state of given Hamiltonian H by applying the following non-unitary operator $\exp(-\beta H)$ (where $\beta$ = $it$). Let $\ket{\psi_{ground}}$ denote the ground state of H and $\ket{\phi (0)}$ be some given random state. The key ingredient of QITE) is the following property:
\begin{align}
    \ket{\psi_{ground}} = \lim_{\beta \rightarrow \infty} \frac{ \ket{\phi(\beta)}} { \sqrt{ \braket{\phi(\beta)| \phi(\beta) }  }},
\end{align}
where $\ket{\phi(\beta)} = \exp(-\beta H) \ket{\phi (0)}$. Therefore, one can think of this QITE method as the continuous version of the classical power method where ideally, we apply for a very long time $t$, e.g., $t \rightarrow \infty$. As described in~\cite{motta2020determining}, the QITE eventually discretizes the operation $\exp(-\beta H)$ with Trotter decomposition  and finds a corresponding effective Hamiltonian so that its real-time evolution approximates a small time step of the imaginary time evolution, via solving a corresponding linear system from measurements. Meanwhile, in our outlined approach, we make use of the Hadamard test procedure to estimate the eigenvalue.

\subsection{Largest few eigenvalues}
\label{sec: largest}
In Ref.~\cite{motta2020determining}, a quantum Lanczos algorithm was proposed to compute ground and excited states of a Hamiltonian $H$, based on  analogously quantum imaginary time evolution $e^{-H\Delta \tau}$ (which is in turn approximated by a unitary evolution under some other Hamiltonian) and the classical Lanczos algorithm. In the classical Lanczos algorithm, the Krylov subspace is built in powers of $H$, whereas in the particular quantum implementation by~\cite{motta2020determining}, the Krylov subspace is built in powers of $\exp(-\Delta\tau H)$. The two approaches (of $H$ or of $e^{-\Delta \tau H}$) become identical in the limit $\Delta\tau \rightarrow 0$. Moreover, in the Lanczos algorithm, a crucial aspect is to compute the following `inner' products $v_j^{*T}v_k$ and $v_j^{*T}Hv_k$, where $v_k$ is some vector with the unit norm in the subspace. In this regard, the procedure outlined in our work is building up the Krylov subspace in terms of a series of quantum states. It can then be  employed to compute the largest few eigenvalues in magnitude by using quantum computers to compute the $H$ matrix in the Krylov subspace and the overlaps among the vectors $H^m x_0$ and $H^n x_0$. We then classically diagonalize the resultant eigenvalue problem in the Krylov subspace, yielding the largest few eigenvalues (in magnitude). We remark that the main algorithm outlined in our work has intermediate measurement, which was used to ensure state normalization. Therefore, it cannot be directly used to estimate the term of the form $(H^m x_0)^{*T} H (H^m x_0)$. But in appendix \ref{sec: krylov}, we show that a minor modification of our algorithm can achieve such a goal of estimating relevant overlaps and matrix elements properly.  

Moreover, by combining the shifting and the choice of applying $H$ or $H^{-1}$, we can generalize our algorithm to obtain a few largest or smallest eigenvalues  either in magnitude or in their algebraic values.  Such capability can potentially lead to more efficient access to, e.g., the ground-state energy and those of a few low-lying excited states, as many physics models possess Hamiltonians whose matrices are spare. \\

\subsection{Classically Enhanced Initialization}
As we have established in Thm.~\ref{thm: quantumrunningtime}, w.r.t.  the dimension $n$, the speedup of our quantum algorithm could be improved from nearly quadratic to exponential, if $x_0$ is already close to the appropriate eigenvector. Quantumly, in order to execute the power method, one could generate $\ket{x_0}$ by choosing a circuit $U$ consisting of rotational gates with tunable parameters and randomly initializing them. The aforementioned $U$ is a variational quantum circuit, which is an extremely popular model for exploring current noisy devices, with major applications in machine learning~\cite{mitarai2018quantum, schuld2020circuit, schuld2018supervised, havlivcek2019supervised}. Thus, it is  interesting to draw the relevance of our quantum algorithm with the emergence of near-term quantum computers. 

If we wish to obtain exponential speedup w.r.t $n$, we need $\ket{x_0}$ to be close to the eigenvector $\ket{E_n}$ that corresponds to the largest eigenvalue. It is quite straightforward to see that, for a given $A$ in our assumption (see Thm. 1 and Eq.~\ref{thm: quantumrunningtime}), the value,
\begin{align}
    \alpha = \bra{\phi} A \ket{\phi}
\end{align}
is maximized only when $\ket{\phi}$ is exactly $\ket{E_n}$. If $\ket{\phi}$ is generated by a variational quantum circuit from trivial state $\ket{\bf 0}$, then we remark that the problem of maximizing/minimizing $\alpha$ is very standard within the context of the variational quantum algorithm (VQA)~\cite{cerezo2021cost}. A very popular hybrid method currently being widely used is the gradient descent, where a cost function $C$ is defined in terms of $\alpha$, then a combination of quantum and classical computer is employed to estimate the gradient and carry out the optimization, following by an update of parameters in  the quantum circuit. In reality, we might not obtain exactly the state $\ket{E_n}$, but the above optimization process can drive the initial state, say, $\ket{\phi} \equiv \ket{x_0}$ closer to $\ket{E_n}$. Thus, in principle, the heuristic VQA method can enhance the quantum power method by increasing the quality of the initial seed $\ket{x_0}$, before executing the matrix multiplication. However, this way of enhancement does not seem friendly to near-term devices, as we need the ability to perform multiplication by $A$.
One might suspect that, given the ability to multiply $A$, if one can maximize $\alpha$, then we already achieve the eigenvector and the largest eigenvalue purely by the variational method. Then what should be the point of the power method? We remind that while the variational method seems appealing and easy to implement, it is still a heuristic method with an arguably unknown performance guarantee. For instance, we do not know how many iterations and the circuit depth are necessary to obtain the eigenvalue with a desired accuracy $\epsilon$. Additionally, as pointed out by many works, such as~\cite{cerezo2021cost, wang2021noise}, training variational circuits can encounter the so-called barren plateau issue, i.e., vanishing gradients, which might severely affect the VQA method. At the same time, as we elaborate in Appendix~\ref{sec: erroranalysis}, the power method (both classical and quantum) is a rigorous tool with guaranteed performance. Thus, as we pointed out here, a combination of both VQA and the power method can be very useful.  \\

\section{Conclusion}\label{sec:conclusions}
Inspired by many developments in quantum algorithms, we have outlined an efficient quantum algorithm for estimating the largest absolute eigenvalue of a sparse Hermitian matrix. Possible extensions to non-sparse matrices and to solving related eigenvalues problems were also discussed. The core routines of our algorithm are the consecutive application of the adaptive version of HHL algorithm~\cite{harrow2009quantum, wiebe2012quantum} and the Hadamard test, both of which have proved to be useful in many contexts, such as in Refs.~\cite{lloyd2013quantum, motta2020determining, wiebe2012quantum}. Our work thus suggests an efficient quantum approach to a particular computational problem, i.e., solving the eigenvalue problem, for which  classical algorithms require polynomial time with respect to the size of the input matrix.  Identifying difficult computational problems and finding efficient solutions in both classical and quantum contexts has  been one major driver to improve computation capability and extend its applications. Our work has contributed another small step to enrich the existing quantum algorithm pool and can be used as a subroutine (just as the HHL algorithm) in other quantum algorithms or future applications.  \\

\medskip
\noindent {\bf Acknowledgements}.
This work was supported in part  by the U. S. Department of Energy, Office of Science, National Quantum Information Science Research Centers, Co-design Center for Quantum Advantage (C2QA)
under contract number DE-SC0012704, in particular, for the design of the main algorithm, and  by the National Science Foundation under Grant No. PHY
1915165, in particular, for the extension of the algorithm and its application to physical Hamiltonians.

\bibliography{ref.bib}{}

\begin{thebibliography}{10}

\bibitem{feynman2018simulating}
Richard~P Feynman.
\newblock Simulating physics with computers.
\newblock In {\em Feynman and computation}, pages 133--153. CRC Press, 2018.

\bibitem{deutsch1985quantum}
David Deutsch.
\newblock Quantum theory, the church--turing principle and the universal
  quantum computer.
\newblock {\em Proceedings of the Royal Society of London. A. Mathematical and
  Physical Sciences}, 400(1818):97--117, 1985.

\bibitem{grover1996fast}
Lov~K Grover.
\newblock A fast quantum mechanical algorithm for database search.
\newblock In {\em Proceedings of the twenty-eighth annual ACM symposium on
  Theory of computing}, pages 212--219, 1996.

\bibitem{shor1999polynomial}
Peter~W Shor.
\newblock Polynomial-time algorithms for prime factorization and discrete
  logarithms on a quantum computer.
\newblock {\em SIAM review}, 41(2):303--332, 1999.

\bibitem{childs2010quantum}
Andrew~M Childs and Wim Van~Dam.
\newblock Quantum algorithms for algebraic problems.
\newblock {\em Reviews of Modern Physics}, 82(1):1, 2010.

\bibitem{berry2007efficient}
Dominic~W Berry, Graeme Ahokas, Richard Cleve, and Barry~C Sanders.
\newblock Efficient quantum algorithms for simulating sparse hamiltonians.
\newblock {\em Communications in Mathematical Physics}, 270(2):359--371, 2007.

\bibitem{harrow2009quantum}
Aram~W Harrow, Avinatan Hassidim, and Seth Lloyd.
\newblock Quantum algorithm for linear systems of equations.
\newblock {\em Physical review letters}, 103(15):150502, 2009.

\bibitem{childs2017quantum}
Andrew~M Childs, Robin Kothari, and Rolando~D Somma.
\newblock Quantum algorithm for systems of linear equations with exponentially
  improved dependence on precision.
\newblock {\em SIAM Journal on Computing}, 46(6):1920--1950, 2017.

\bibitem{lloyd2013quantum}
Seth Lloyd, Masoud Mohseni, and Patrick Rebentrost.
\newblock Quantum algorithms for supervised and unsupervised machine learning.
\newblock {\em arXiv preprint arXiv:1307.0411}, 2013.

\bibitem{lloyd2014quantum}
Seth Lloyd, Masoud Mohseni, and Patrick Rebentrost.
\newblock Quantum principal component analysis.
\newblock {\em Nature Physics}, 10(9):631--633, 2014.

\bibitem{lloyd2016quantum}
Seth Lloyd, Silvano Garnerone, and Paolo Zanardi.
\newblock Quantum algorithms for topological and geometric analysis of data.
\newblock {\em Nature communications}, 7(1):1--7, 2016.

\bibitem{lloyd2020quantum}
Seth Lloyd, Maria Schuld, Aroosa Ijaz, Josh Izaac, and Nathan Killoran.
\newblock Quantum embeddings for machine learning.
\newblock {\em arXiv preprint arXiv:2001.03622}, 2020.

\bibitem{schuld2018supervised}
Maria Schuld and Francesco Petruccione.
\newblock {\em Supervised learning with quantum computers}, volume~17.
\newblock Springer, 2018.

\bibitem{mitarai2018quantum}
Kosuke Mitarai, Makoto Negoro, Masahiro Kitagawa, and Keisuke Fujii.
\newblock Quantum circuit learning.
\newblock {\em Physical Review A}, 98(3):032309, 2018.

\bibitem{nghiem2022constant}
Nhat~A Nghiem, Xianfeng~David Gu, and Tzu-Chieh Wei.
\newblock Constant-time quantum algorithm for homology detection in closed
  curves.
\newblock {\em arXiv preprint arXiv:2209.12298}, 2022.

\bibitem{clader2013preconditioned}
B.~D. Clader, B.~C. Jacobs, and C.~R. Sprouse.
\newblock Preconditioned quantum linear system algorithm.
\newblock {\em Phys. Rev. Lett.}, 110:250504, Jun 2013.

\bibitem{wossnig2018quantum}
Leonard Wossnig, Zhikuan Zhao, and Anupam Prakash.
\newblock Quantum linear system algorithm for dense matrices.
\newblock {\em Physical review letters}, 120(5):050502, 2018.

\bibitem{wiebe2012quantum}
Nathan Wiebe, Daniel Braun, and Seth Lloyd.
\newblock Quantum algorithm for data fitting.
\newblock {\em Physical review letters}, 109(5):050505, 2012.

\bibitem{nielsen2002quantum}
Michael~A Nielsen and Isaac Chuang.
\newblock Quantum computation and quantum information, 2002.

\bibitem{chu1993multivariate}
Moody~T Chu and J~Loren Watterson.
\newblock On a multivariate eigenvalue problem, part i: Algebraic theory and a
  power method.
\newblock {\em SIAM Journal on scientific computing}, 14(5):1089--1106, 1993.

\bibitem{parlett1982estimating}
Beresford~N Parlett, H~Simon, and LM~Stringer.
\newblock On estimating the largest eigenvalue with the lanczos algorithm.
\newblock {\em Mathematics of computation}, 38(157):153--165, 1982.

\bibitem{o1979estimating}
Dianne~P O’Leary, GW~Stewart, and James~S Vandergraft.
\newblock Estimating the largest eigenvalue of a positive definite matrix.
\newblock {\em Mathematics of Computation}, 33(148):1289--1292, 1979.

\bibitem{wiebe2014quantum}
Nathan Wiebe, Ashish Kapoor, and Krysta Svore.
\newblock Quantum algorithms for nearest-neighbor methods for supervised and
  unsupervised learning.
\newblock {\em arXiv preprint arXiv:1401.2142}, 2014.

\bibitem{brassard2002quantum}
Gilles Brassard, Peter Hoyer, Michele Mosca, and Alain Tapp.
\newblock Quantum amplitude amplification and estimation.
\newblock {\em Contemporary Mathematics}, 305:53--74, 2002.

\bibitem{yuster2005fast}
Raphael Yuster and Uri Zwick.
\newblock Fast sparse matrix multiplication.
\newblock {\em ACM Transactions On Algorithms (TALG)}, 1(1):2--13, 2005.

\bibitem{tao2011random}
Terence Tao and Van Vu.
\newblock Random matrices: universality of local eigenvalue statistics.
\newblock {\em Acta mathematica}, 206(1):127--204, 2011.

\bibitem{kitaev1995quantum}
A~Yu Kitaev.
\newblock Quantum measurements and the abelian stabilizer problem.
\newblock {\em arXiv preprint quant-ph/9511026}, 1995.

\bibitem{kerenidis2016quantum}
Iordanis Kerenidis and Anupam Prakash.
\newblock Quantum recommendation systems.
\newblock {\em arXiv preprint arXiv:1603.08675}, 2016.

\bibitem{yu2017finding}
Xiongjie Yu, David Pekker, and Bryan~K. Clark.
\newblock Finding matrix product state representations of highly excited
  eigenstates of many-body localized hamiltonians.
\newblock {\em Phys. Rev. Lett.}, 118:017201, Jan 2017.

\bibitem{wang2022quantum}
Guoming Wang, Daniel Stilck-Fran{\c{c}}a, Ruizhe Zhang, Shuchen Zhu, and
  Peter~D Johnson.
\newblock Quantum algorithm for ground state energy estimation using circuit
  depth with exponentially improved dependence on precision.
\newblock {\em arXiv preprint arXiv:2209.06811}, 2022.

\bibitem{lin2022Heisenberg}
Lin Lin and Yu~Tong.
\newblock Heisenberg-limited ground-state energy estimation for early
  fault-tolerant quantum computers.
\newblock {\em PRX Quantum}, 3:010318, Feb 2022.

\bibitem{motta2020determining}
Mario Motta, Chong Sun, Adrian~TK Tan, Matthew~J O’Rourke, Erika Ye, Austin~J
  Minnich, Fernando~GSL Brand{\~a}o, and Garnet~Kin Chan.
\newblock Determining eigenstates and thermal states on a quantum computer
  using quantum imaginary time evolution.
\newblock {\em Nature Physics}, 16(2):205--210, 2020.

\bibitem{schuld2020circuit}
Maria Schuld, Alex Bocharov, Krysta~M Svore, and Nathan Wiebe.
\newblock Circuit-centric quantum classifiers.
\newblock {\em Physical Review A}, 101(3):032308, 2020.

\bibitem{havlivcek2019supervised}
Vojt{\v{e}}ch Havl{\'\i}{\v{c}}ek, Antonio~D C{\'o}rcoles, Kristan Temme,
  Aram~W Harrow, Abhinav Kandala, Jerry~M Chow, and Jay~M Gambetta.
\newblock Supervised learning with quantum-enhanced feature spaces.
\newblock {\em Nature}, 567(7747):209--212, 2019.

\bibitem{cerezo2021cost}
Marco Cerezo, Akira Sone, Tyler Volkoff, Lukasz Cincio, and Patrick~J Coles.
\newblock Cost function dependent barren plateaus in shallow parametrized
  quantum circuits.
\newblock {\em Nature communications}, 12(1):1791, 2021.

\bibitem{wang2021noise}
Samson Wang, Enrico Fontana, Marco Cerezo, Kunal Sharma, Akira Sone, Lukasz
  Cincio, and Patrick~J Coles.
\newblock Noise-induced barren plateaus in variational quantum algorithms.
\newblock {\em Nature communications}, 12(1):6961, 2021.

\bibitem{friedman1998error}
Joel Friedman.
\newblock Error bounds on the power method for determining the largest
  eigenvalue of a symmetric, positive definite matrix.
\newblock {\em Linear algebra and its applications}, 280(2-3):199--216, 1998.

\bibitem{schuld2014quest}
Maria Schuld, Ilya Sinayskiy, and Francesco Petruccione.
\newblock The quest for a quantum neural network.
\newblock {\em Quantum Information Processing}, 13(11):2567--2586, 2014.

\bibitem{schuld2019evaluating}
Maria Schuld, Ville Bergholm, Christian Gogolin, Josh Izaac, and Nathan
  Killoran.
\newblock Evaluating analytic gradients on quantum hardware.
\newblock {\em Physical Review A}, 99(3):032331, 2019.

\bibitem{prakash2014quantum}
Anupam Prakash.
\newblock {\em Quantum algorithms for linear algebra and machine learning}.
\newblock University of California, Berkeley, 2014.

\bibitem{plesch2011quantum}
Martin Plesch and {\v{C}}aslav Brukner.
\newblock Quantum-state preparation with universal gate decompositions.
\newblock {\em Physical Review A}, 83(3):032302, 2011.

\end{thebibliography}
\bibliographystyle{unsrt}

\clearpage
\newpage
\onecolumngrid

\appendix
\section{Analysis of Error}
\label{sec: erroranalysis}
We remark that the original classical algorithm described in Sec.~\ref{sec: classicalalgorithm} only produces an approximation to $\lambda_n$. Here we provide a more detailed analysis of the error. It is clear that the error vanishes in the limit $k \rightarrow \infty$. \\

For simplicity of the presentation, we assume $A$ to be real and symmetric, but the extension to the Hermitian case is straightforward. 
Let $x_k = A^k x_0$ and $\bar{x}_k$ be: 
$$ \bar{x}_k \equiv \lambda_n^k c_n E_n. $$
We note that, the \textit{exact} eigenvalue $\lambda_n$ is: 
\begin{align}
    \lambda_n = \frac{ \Bar{x}_{k+1} \cdot \bar{x}_k }{ \bar{x}_k\cdot \bar{x}_k }.
\end{align}
The approximated eigenvalue is:
\begin{align}
    \bar{\lambda}_n = \frac{Ax_k \cdot x_k}{x_k \cdot x_k}.
\end{align}
From Eqn.~(\ref{eqn: xk}), we have: 
\begin{align}
A x_k = \lambda_n^{k+1} \Big( c_1\frac{\lambda_1^{k+1}}{\lambda_n^{k+1}}E_1 + c_2\frac{\lambda_2^{k+1}}{\lambda_n^{k+1}}E_2 + \cdots + c_nE_n \Big).
\end{align}
By using the property that $\{E_i\}_1^n$ are orthonormal, we have:
\begin{align}
    Ax_k \cdot x_k = \lambda_n^{k+1}\lambda_n^k \Big( c_1^2  \frac{\lambda_1^{k+1}}{\lambda_n^{k+1}}\frac{\lambda_1^{k}}{\lambda_n^{k}}  + c_2^2 \frac{\lambda_2^{k+1}}{\lambda_n^{k+1}}\frac{\lambda_2^{k}}{\lambda_n^{k}}  + \cdots + c_n^2   \Big).
\end{align}
We also have:
\begin{align}
    x_k\cdot x_k = \lambda_n^{2k} \Big(c_1^2 \frac{\lambda_1^{2k}}{\lambda_n^{2k}} + c_2^2 \frac{\lambda_2^{2k}}{\lambda_n^{2k}} \cdots +   c_n^2  \Big). 
\end{align}

Straightforward computations yield:
\begin{align}
    \bar{x}_k\cdot \bar{x}_k = \lambda_n^{2k} c_n^2, \\
    \bar{x}_{k+1} \cdot \bar{x}_k = \lambda_n^{k+1}\lambda_n^{k} c_n^2.
\end{align}

Now we aim to compute the absolute ratio between the approximated eigenvalue and the exact eigenvalue: 
\begin{align}
    \Big|\frac{\bar{\lambda}_n}{\lambda_n}\Big| &= \Big|\frac{Ax_k\cdot x_k}{\bar{x}_{k+1} \cdot \bar{x}_k } \cdot \frac{ \bar{x}_k\cdot \bar{x}_k }{ x_k \cdot x_k }\Big|.
\end{align}
A simple substitution yields
\begin{align}
    \frac{ \bar{x}_k\cdot \bar{x}_k }{ x_k \cdot x_k } &= \frac{1}{1 + \sum_i (\frac{c_i}{c_n})^2 \cdot ( \frac{\lambda_i}{\lambda_n})^{2k} } < 1.
\end{align}
Therefore, we arrive at
$$  \Big|\frac{\bar{\lambda}_n}{\lambda_n}\Big| < \Big| \frac{Ax_k\cdot x_k}{\bar{x}_{k+1} \cdot \bar{x}_k }  \Big|. $$

Another round of substitution yields: 
\begin{align}
    \frac{Ax_k\cdot x_k}{\bar{x}_{k+1} \cdot \bar{x}_k } &= 1 + \sum_{i=1}^{n-1} \frac{c_i^2}{c_n^2} \frac{\lambda_i^k}{\lambda_n^k} \frac{\lambda_i^{k+1}}{\lambda_n^{k+1}}.
    \label{eqn: ratio}
\end{align}

We then make use of the following conditions: \\

$\bullet$ Since $\lambda_i/\lambda_n< $1 for all $i<n$ then 
$$\frac{\lambda_i^k}{\lambda_n^k} \frac{\lambda_i^{k+1}}{\lambda_n^{k+1}} < \frac{\lambda_i^{2k}}{\lambda_n^{2k}}.  $$

$\bullet$ $0 \leq \lambda_1 \leq \lambda_2 \leq \cdots < \lambda_n$ then for all $i<n$:
$$ \frac{\lambda_i}{\lambda_n} \leq \frac{\lambda_{n-1}}{\lambda_n} = p <1. $$

$\bullet$ All the components $c_1,c_2, ..., c_n$ are bounded above, therefore, the ratio $c_i/c_n$ is bounded above by some value $K$ for all $i$. 

Therefore, the ratio in Eqn.~\ref{eqn: ratio} is upper bounded,
\begin{align}
     \frac{Ax_k\cdot x_k}{\bar{x}_{k+1} \cdot \bar{x}_k } < 1 + (n-1)\cdot K^2 \cdot p^{2k}.
\end{align}
Altogether, we have:
\begin{align}
     \Big|\frac{\bar{\lambda}_n}{\lambda_n}\Big| < 1 + (n-1)\cdot K^2 \cdot p^{2k}.
\end{align}

In order to estimate the eigenvalue $\lambda_n$ to some multiplicative error $\delta$, we expect:
\begin{align}
    \Big|\frac{\bar{\lambda}_n}{\lambda_n}\Big| < 1 + \delta.
\end{align}

We thus require that 
$$ (n-1)\cdot K^2 \cdot p^{2k} < \delta, $$
which is satisfied if 
$$ k \in \Omega\big( \log(1/\delta) + \log ((n-1)K ) )/ \log(1/p) \big).$$

Note the above lower bound is the result of a very elementary analysis of the power method. We can easily see that if, by any mean, $K \approx 1/\sqrt{n-1}$, then it will cancel out the $(n-1)$ term and leave us a better condition: $p^{2k} < \delta$, which implies that the lower bound of $k$ is much smaller, i.e, $k \in \Omega( log(1/\delta)/\log(1/p))$. One may ask, when will we have such a nice condition $K \approx 1/\sqrt{n-1}$? Recall that $K$ is an upper bound of $\{c_i/c_n\}$ and all $\{c_i\}$ are coefficients of $x_0$, therefore, its value depends on how the initial vector $x_0$ is randomly chosen. With some luck, if $x_0$ is relatively close to the largest eigenvector $E_n$, which means that $c_n$ is considerably larger than all other coefficients $c_i$, we can have such a nice value of $K$, which yields a better lower bound on $k$ -- the number of required iterations. It is very interesting to ask if we can control such a process, i.e., choosing $x_0$ in an efficient way, by some low-cost method. \\

In general, $x_0$ is random, therefore, we can expect to have some probabilistic analysis of the error. Fortunately, this issue was addressed in \cite{friedman1998error}. The analysis provided in \cite{friedman1998error} is rigorous. Roughly speaking, it yields the following statement:
\begin{lemma}[Error Bound]
\label{lemma: errbound}
Let $\gamma >0$ and $\frac{\lambda_{n-1}}{\lambda_n} = p$ as above. Then with probability $\geq 1 - 2\gamma$, we have the following bound: 
\begin{align}
    \Big|\frac{\bar{\lambda}_n}{\lambda_n}\Big| < 1 + \mathcal{O}( \gamma^{-2} \sqrt{n} p^{2k-2}  ).
\end{align}
\end{lemma}

Therefore, in order for the power method to achieve a multiplicative error $\delta$, using the above result, with a certain probability ($\geq 1-2\gamma$), we have a better condition:
\begin{align}
    \gamma^{-2} \sqrt{n} p^{2k-2} < \delta,
\end{align}
which yields:
$$ k \in \Omega(( \log(1/\delta) + \log \sqrt{n})/ \log(1/p) ). $$

Compared to the previous elementary analysis, the bound given in \cite{friedman1998error} gives an improvement over the dimension n. Now we discuss an issue that might affect the overall performance of the quantum power method. We remind that the classical power method begins with a randomized vector $x_0$. In the quantum setting, this vector could be generated by some random quantum circuit. In general, the form of such a vector is a product state, therefore, it cannot be as purely random as we expect from a classical setting. As a result, the above lemma \ref{lemma: errbound} might not yield the same implication in the quantum setting. For example, since we cannot generate an arbitrary vector, the probability might not be as high as claimed in Lemma~\ref{lemma: errbound}. We wish to point out a solution to this issue, inspired by the amplitude encoding method, which is very popular in the quantum machine learning context~\cite{schuld2014quest, schuld2018supervised, schuld2019evaluating}. The task is that, if we are given a vector $\bf{x} \equiv \{x_1,x_2, ..., x_n \}$ with all known entries (assumed to be normalized for simplicity), and we wish to generate a quantum state $\ket{x} = \sum_{i=1}^n x_i \ket{i}$ that corresponds to $\bf{x}$. Then a method outlined in~\cite{prakash2014quantum} can be used to approximately generate such a state. We summarize the result as follows:

\begin{lemma}~\cite{prakash2014quantum,plesch2011quantum}
\label{lemma: ampenc}
A quantum state $\ket{B} \in R^N$ with $b$ non-zero entries can be approximately prepared using a circuit of size $\mathcal{O}(b)$ and depth $\mathcal{O}(log(b))$.
\end{lemma}

It is worth mentioning that efficient quantum circuits, e.g., circuits of polynomial size, cannot reach all states in the Hilbert space. Therefore, we do not expect that the above amplitude encoding method can be used to generate arbitrary states. The above lemma also reveals that in general, only an approximation of given state $\ket{B}$ is produced. This issue does not affect the quantum power method, as what exactly we want is not the exact state, but the \textit{randomness} of the state. Our solution first proceeds by classically randomizing a vector, then producing an approximation to this vector (all entries are known) using the amplitude encoding method. In this way, the initial quantum state \textit{mimics} the randomly initialized vector.

\section{Details of the HHL-like approach for multiplication}
\label{sec: HHL}
In this section, we explicitly prove the lemma \ref{lemma: matrixapp}. \\

Let the initial state $\ket{x_0}$ be given, and its decomposition in the basis consisting of eigenvectors of A, $\{ \ket{u_i}\}_{i=1}^N$ (with corresponding eigenvalues $\{\lambda_i \}_{i=1}^N$ ) as follows:
$$ \ket{x_0} = \sum_{i=1}^N \beta_i \ket{u_i}. $$
\\
Following HHL,  we first prepare  the state: 
\begin{align}
  U_0 \ket{0} \otimes \ket{x_0} =  \ket{\Phi_0} = \sqrt{\frac{2}{T}} \sum_{\tau = 0}^{T-1} \sin( \frac{\pi(\tau+1/2)}{T})\ket{\tau} \otimes \ket{x_0}.
\end{align}
We then apply the following controlled unitary 
\begin{align}
    U_T = \sum_{\tau=0}^T \ket{\tau}\bra{\tau} \otimes \exp(\frac{-iA\tau t_0}{T}).
\end{align}
to the above state, followed by a quantum Fourier transform over the first register. We approximately arrive at  (we will elaborate on the approximation later)
\begin{align}
    \ket{\phi} = \sum_{i=1}^N \beta_i \ket{\lambda_i}\ket{u_i}.
    \label{eqn: state}
\end{align}
Next, we  append an ancilla initialized in $\ket{0}$ and perform the following  rotation on it controlled on the phase register:
\begin{align*}
    \ket{\lambda_i}\ket{0} \rightarrow  \ket{\lambda_i}\big( C\lambda_i \ket{0} + \sqrt{1 - C^2\lambda_i^2}\ket{1}\big),
\end{align*}
where the constant $C$ secures the normalization condition and is chosen  so that  
\begin{align*}
     C\lambda_{max} \leq 1.
\end{align*}
Since we have the condition: $1/\kappa \leq \lambda_i \leq 1$ for all i, therefore, the condition for $C$ can be further simplified as 
\begin{align*}
    C \leq \kappa.
\end{align*}

After the above step, the state $\ket{\phi}$ is transformed to:
\begin{align*}
    \ket{\phi_1} = \sum_{i=1}^N \beta_i \ket{\lambda_i}\ket{u_i}\big( C\lambda_i\ket{0} + \sqrt{1 - C^2\lambda_i^2}\ket{1}   \big).
\end{align*}
We note that, our desired vector $A\ket{x_0}$ is 
\begin{align*}
    A\ket{x_0} = \sum_{i=1}^N \lambda_i \beta_i \ket{u_i}.
\end{align*}
Therefore, the state $\ket{\phi_1}$, after we uncompute the phase register, which could be done by running $U_T^\dagger$ and $U_0^\dagger$, would approximately become 
\begin{align}
 \label{eqn: firstroundstate}
    \ket{\phi_1} = C\Vert A\ket{x_0}\!\Vert\ket{A\ket{x_0}} \ket{0} + \ket{G_1}\ket{1},
\end{align}
where $\ket{G_1}$ (not normalized properly), as mentioned earlier, is some unimportant state; and $\Vert A\ket{x_0}\Vert$ is the length ($l_2$-norm) of the vector $A\ket{x_0}$.  We have emphasized previously that we only obtained an approximation to Eqn.~(\ref{eqn: state}) (and hence, the state Eqn.~(\ref{eqn: firstroundstate}) as well), which is due to the finite-bit encoding of the phase value~\cite{harrow2009quantum}. A thorough analysis has been provided in~\cite{harrow2009quantum}, showing that if 
%\begin{align*}
    $t_0 =\mathcal{O}(\kappa/\epsilon)$,
%\end{align*}
then the total final error is $\epsilon$. By the final error, we mean that the state $\ket{A\ket{x_0}}$ in Eqn.~(\ref{eqn: firstroundstate}) is $\epsilon$-close to the actual solution. Throughout the discussions below, we will keep using the notation $\ket{A\ket{x_0}}$ to denote the approximated state with the understanding that it is just  $\epsilon$-close to the real solution (in the main text, we actually distinguish the two).

Instead of measuring the ancilla and trying to project to the state $|Ax_0\rangle$, as did in the HHL algorithm, our next step is to apply $A$ again to the state $\ket{A\ket{x_0}}$, or more precisely, the vector $Ax_0$ itself. (We note that a simpler alternative way is to continue to apply the controlled rotation on an additional ancilla without undoing the phase estimation; see Appendix~\ref{app:Improved} below.)  Again, let the decomposition of $\ket{A\ket{x_0}}$ in the basis $\{\ket{u_i} \}_{i=1}^N$ as,
\begin{align*}
    \ket{A\ket{x_0}} = \sum_{i=1}^N \beta'_i \ket{u_i}.
\end{align*}
It is  straightforward to see that since $A\ket{x_0} = \sum_{i=1}^N \lambda_i\beta_i \ket{u_i}$, then the coefficient $\beta'_i$ in the above state is just 
\begin{align*}
    \beta'_i = \frac{\lambda_i \beta_i}{\Vert A\ket{x_0}\Vert} = \frac{ \lambda_i\beta_i}{\sqrt{\sum_{i=1}^N |\lambda_i\beta_i|^2}}.
    \label{eqn: betaiprime}
\end{align*}
 Following the same procedure as we did for the application of $A$ to $\ket{x_0}$, we first prepare the state $\ket{\Phi_0}$ and append another ancilla $\ket{0}$, then we  obtain approximately the following state
\begin{align}
    \ket{\phi_2} &= C\Vert A\ket{x_0}\Vert \Big( \sum_{i=1}^N \beta'_i \ket{\lambda_i}\ket{u_i}   \Big) \ket{0}\ket{0} +  U_T\ket{\Phi_0}\ket{G_1}\ket{1}\ket{0}.
\end{align}
We take a specific look at the first part of the above state (we ignore everything entangled with $\ket{G_1}$ as they are not important). At this step, we would rotate the ancilla $\ket{0}$ controlled by the phase register. However, a subtle point here is that this time, the rotation needs to be controlled further by the qubit $\ket{0}$ that is resulted from the previous application of $A$ (note that there are two ancillary qubits $\ket{0}\ket{0}$ in the above state). With such detail in mind, we will obtain the following state:
\begin{align}
    C\Vert A\ket{x_0}\Vert \Big( \sum_{i=1}^N \beta'_i \ket{\lambda_i}\ket{u_i}\ket{0} ( C\lambda_i \ket{0} + \sqrt{1-C^2\lambda^2_i}\ket{1}   )   \Big) + U_T\ket{\Phi_0}\ket{G_1}\ket{1,0}.
\end{align}
Now we uncompute the phase register (and omit it afterward) to obtain:
\begin{align}
    C^2 \Vert A\ket{x_0}\Vert \sum_{i=1}^N \beta'_i \lambda_i \ket{u_i}\ket{00} + \ket{G_2}\ket{01} + \ket{G_1}\ket{1}\ket{0}.
    \label{eqn: seconditeration}
\end{align}
We remark that since $\{\beta'_i\}_{i=1}^N$ are the coefficients of the decomposition of the \textit{normalized state} $\ket{A\ket{x_0}}$ in the basis $\{\ket{u_i}\}_{i=1}^N$, then multiplying with the factor $\Vert A\ket{x_0}\Vert$ (which is the actual length of the original vector $A\ket{x_0}$) would yield the coefficient of the original vector $A\ket{x_0}$ in the same basis. From Eqn.~\ref{eqn: betaiprime}, it directly leads to: 
\begin{align}
    \Vert A\ket{x_0}\Vert\sum_{i=1}^N \beta'_i \lambda_i \ket{u_i} = A(A\ket{x_0}) = A^2\ket{x_0}.
    \label{eqn: multiplying}
\end{align}
Note that this is a vector that is not normalized. Hence, the Eqn.~(\ref{eqn: seconditeration}) can be written simpler as
\begin{align}
    C^2 \Vert A^2\ket{x_0}\Vert \ket{A^2\ket{x_0}} \ket{00} + \ket{G_2}\ket{01} + \ket{G_1}\ket{10}.
\end{align}
Iterating the procedure $k$ times, we eventually obtain what we got in Lemma \ref{lemma: matrixapp}. \\

As a final remark, the property that we have used in Eqn.~(\ref{eqn: multiplying}) is indeed very useful. After each iteration, there is a length factor (such as $\Vert A\ket{x_0}\Vert$ in the above) but eventually, it will get `absorbed' in the next iteration, which is very convenient as we only care about the final vector $A^k\ket{x_0}$ in order to estimate $\lambda_{max}$. 

\section{Improved Approach}
\label{app:Improved}
The phase rotation procedure can be improved by almost a factor of $k$--the number of iterations. The application of $A$ on $\ket{x_0}$ can be written explicitly as:
\begin{align}
    A\ket{x_0} = \sum_{i=1}^N \lambda_i \beta_i \ket{u_i}.
\end{align}
Likewise, another application of $A$ yields:
\begin{align}
    A^2 \ket{x_0} = \sum_{i=1}^N \lambda^2_i \beta_i \ket{u_i}.
\end{align}
After $k$ times, we have:
\begin{align}
    A^k \ket{x_0}= \sum_{i=1}^N \lambda^k_i \beta_i \ket{u_i},
    \label{eqn: kapplication}
\end{align}
as we have described earlier.

As we have emphasized, the key step in our quantum algorithm is the execution of $\exp(-iAt)$ in QPE to extract the eigenvalues of $A$, and perform the controlled rotation. As shown in~\cite{harrow2009quantum, wiebe2012quantum}, after running QPE and controlled rotation in the first iteration, we would have the following state: 
\begin{align}
    \sum_{i=1}^N \beta_i \ket{\lambda_i}\ket{u_i}\Big( C\lambda_i\ket{0}+ \sqrt{1-C^2\lambda^2_i}\ket{1} \Big)
    &= C \sum_{i=1}^N \beta_i \lambda_i \ket{\lambda_i} \ket{u_i}\ket{0} + \ket{G_1}\ket{1}.
\end{align}

Now, instead of uncomputing the phase and repeating the process, we simply append the ancilla $\ket{0}$ and rotate the ancilla controlled by the phase register. In other words, we would have the following rotation: 

\begin{align}
    \ket{\lambda_i} \ket{0} \rightarrow C^k \lambda_i^k \ket{0} + \sqrt{1 -C^{2k} \lambda_i^{2k}}\ket{1}.
\end{align}

Note that for the matrix multiplication purpose, the factor C can be set to 1 given the range of absolute eigenvalues lying between 0 and 1 (see also \cite{wiebe2014quantum}). The above rotation can be achieved by simply noting that the phase $\lambda_i$ is stored digitally, which is very straightforward to apply the arithmetic operation. By doing this way, we do not need to iterate a single application of matrix $A$ $k$ times consecutively. 

\section{Some detail on Krylov Subspace Method}
\label{sec: krylov}
In this section, we show how to estimate the inner product between matrix-multiplied vectors, which is crucial in the Krylov subspace method that we discussed in Section~\ref{sec: largest}. In Section~\ref{sec: largest}, we use $H$, but here we use $A$ to denote the matrix of consideration. 
The basic idea is to get a set of vectors: $x_0$, $x_1=A x_0$, \dots, $x_{m+1}=A^{m}x_0$, which forms a subspace of dimension not greater than $m+1$. Without loss of generalization, we can set $x_0$ to be normalized. Then we need to compute $H_{jk}=x_j^\dagger A x_k$ and $S_{jk}= x_j^\dagger x_j$ and then solve classically the generalized eigenvalue problem $H \psi= E S \psi$. The largest few eigenvalues will be our estimates for the largest few eigenvalues of $A$. We will explain how to compute the two matrices using quantum computers. \\

Recall that we are interested in estimating the following term: $(A^n x_0)^\dagger A (A^m x_0)$ where $x_0$ is some random unit vector and $n,m$ are some integers (note that these $n,m$ here might not the same as those ones above). We use lemma \ref{lemma: improvedmatrix} to create the following state, where we have suppressed a known factor $C^{m,n}$: 
\begin{align}
  U_{A^m}\ket{0}\ket{x_0} =\ket{\Phi_1}  = \ket{0}A^m \ket{x_0} + \ket{1}\ket{Garbage}, \\ 
  U_{A^n} \ket{0} \ket{x_0} = \ket{ \Phi_2} = \ket{0}\ket{A^n}\ket{x_0} + \ket{1}\ket{Garbage}.
\end{align}
Note that in the above, we abuse the use of the notation $\ket{Garbage}$ to denote the unimportant  parts of the state, as they may not be the same. Now we append another ancilla $\ket{0}$ to each of these state to obtain $\ket{0}\ket{\Phi_1}, \ket{0}\ket{\Phi_2}$. Now for state $\ket{0}\ket{\Phi_1}$, we have:
\begin{align}
    \ket{0}\ket{\Phi_1} = \ket{0}\ket{0} A^m \ket{x_0} + \ket{0}\ket{1}\ket{Garbage}.
\end{align}

We use the second ancilla  as a control bit (conditioned on being $\ket{1}$) to flip the first ancilla and  obtain: 
\begin{align}
    \ket{0}\ket{\Phi_1}' = \ket{0}\ket{0} A^m \ket{x_0} + \ket{1}\ket{1}\ket{Garbage}.
\end{align}
We denote the action $U_{A^m}$ plus bit the controlled flip step above as $U_{A^m}'$. 

It is quite straightforward to see that the inner product $\bra{0,\Phi_2} \cdot\ket{0,\Phi_1}' = \bra{x_0} (A^n)^\dagger A^m \ket{x_0}$ up to the known factor $C^{n+m}$. Therefore, we can use the Hadamard test to estimate such overlaps. To do this, we construct a controlled circuit using the above $U_{A^m}'$ and $U_{A^n}$ as well as one additional ancilla qubit,  
$|0\rangle\langle 0|\otimes U_{A^m}'+ |1\rangle\langle 1|\otimes U_{A^n}$, so that it creates a superposition of $(|0\rangle\otimes \ket{0}\ket{\Phi_1^{'}}+|1\rangle \otimes \ket{0} \ket{\Phi_2})/\sqrt{2}$ from an initial  state $|+\rangle\otimes|0\dots0\rangle$. Measuring the ancilla in the $|\!+\!/\!-\rangle$ basis (Pauli X) gives the real part of the overlap. In addition, measuring ancilla in the $|\!+\!i/\!-\!i\rangle$ (Pauli Y) basis gives the imaginary part of the overlap.
In order to estimate such an overlap up to an additive error $\Delta$, it takes $\mathcal{O}(1/\Delta^2)$ repetitions, which could be improved to $\mathcal{O}(1/\Delta)$ using amplitude estimation \cite{brassard2002quantum}.  \\

As straightforward as it seems, the above procedure essentially yields a solution to our Krylov subspace problem, where the goals are: estimating entries of $H$ and $A$. 
Note that the entry $H_{jk}=x_j^\dagger A x_k = \bra{x_0} {A^{j-1}}^\dagger A (A^{k-1}\ket{x_0} ) = \bra{x_0} {A^{j-1}}^\dagger A^k \ket{x_0}$.  By choosing the power $n,m$ appropriately, we can obtain $H_{jk}$ and $S_{jk}$. The last step is solving the generalized eigenvalue problem $H \psi= E S \psi$ via a classical computer.  \\

\end{document}